# The Cosmic Microwave Background: a strange characteristic

A. Dinculescu [1]

**Abstract** The ratio of the self-gravitational energy density of the scattering particles in the universe to the energy density of the scattered photons in the cosmic microwave background (*CMB*) is the same in any volume of space. These two energy densities are equal at a radiation temperature on the order of the present CMB temperature.

_____

e-mail: *adinculescu@cox.net*

One can hardly overestimate the importance of the cosmic microwave background (*CMB*) for the standard cosmological model. What started as an inference of an "effective temperature of space"[1] based on the "absorption lines of the cyanogen radical" in the interstellar clouds[2], surfaced 7 years latter as a prerequisite for the observed "cosmic abundances of the light elements"[3,4], fell into oblivion for almost a quarter of a century and was finally brought to light as "a measurement of excess antenna temperature"[5], proved to be one of the best pieces of evidence for the big bang model of the universe. The existence of the *CMB* not only shows that the universe was once very hot and dense, but allows us to learn more about the origin of galaxies and large scale structures of the universe, and to estimate the values of the basic parameters of the standard cosmological model. It is therefore very important to know as much as possible about the properties and characteristics of this important source of information. With this objective in mind we present here a strange characteristic of the *CMB*, which can be a coincidence or can be the result of a yet unknown property.

Let us consider an arbitrary volume of space of radius $r$ containing a fully ionized, homogenous mixture of matter and thermal radiation. If the photon number density $n_\gamma$ is much larger than the number density $n_e$ of scattering particles (as is the case with the *CMB*) all particles are practically immersed in a bath of thermal radiation. The scattering particles (mostly free electrons) in the radiation bath scatter the incoming photons, and act as sources of radiation of luminosity $L_e$. The collision rate of a free electron with a photon of frequency $\omega$ in the range $d\omega$ is

$$df_e = c\,\sigma_T\,n_\gamma(\omega)\,d\omega \tag{1}$$

where $c$ is the speed of light and $\sigma_T$ is the Thomson cross section. Due to the electric field in the electromagnetic wave the electron accelerates and emits radiation of the same frequency. The energy radiated in unit time in the above range is

$$dL_e = \hbar\,\omega\,df_e \tag{2}$$

where $h = 2\pi\hbar$ is the Planck constant. Using the Planck distribution

$$n_\gamma(\omega)\,d\omega = \frac{1}{\pi^2 c^3}\,\frac{\omega^3\,d\omega}{e^{\hbar\omega/kT}-1} \tag{3}$$

where $k$ is the Boltzman constant one obtains[6]

$$L_e = c\,\sigma_T \int_0^\infty \frac{\hbar}{\pi^2 c^3}\,\frac{\omega^3\,d\omega}{e^{\hbar\omega/kT}-1} = c\,\sigma_T\,u_\gamma \tag{4}$$

where $u_\gamma = a\,T_\gamma^4$ is the radiation density, $T_\gamma$ is the radiation temperature, and

$$a = \frac{\pi^2}{15}\,\frac{(kT_\gamma)^4}{(\hbar c)^3} \tag{5}$$

is the radiation density constant. With the scattered luminosity per unit volume $L_{ne} = n_e\,\sigma_T\,c\,u_\gamma$ and the escape time of a photon from the volume $t_{esc} = \sigma_T\,n_e\,r^2/c$, the energy density of the scattered radiation inside is $u_* = L_{ne}\,t_{esc} = u_\gamma\,\tau^2$. If one ignores those particles that practically do not play any role in the process (the protons, the neutrons and the unscattered photons), one is left out with a mixture of free electrons and scattered photons that oppose gravitational attraction between them. It is therefore natural to compare the energy density $u_* = u_\gamma\,\tau^2$ of the scattered photons with the self-gravitational energy density

$$u_{eG} = \frac{3}{5} \frac{G M_e^2}{\frac{4\pi}{3} r^4} \tag{6}$$

of the scattering particles. With

$$M_e = \frac{4\pi}{3} n_e m_e r^3 \tag{7}$$

where $m_e$ is the mass of the electron, the two energy densities are equal when

$$a T_\gamma^4 = \frac{4\pi}{5} \frac{G m_e^2}{\sigma_T^2} \tag{8}$$

This corresponds to a radiation temperature practically equal to the present *CMB* temperature[7]. The apparent equality between the self-gravitational energy density of the free electrons and the energy density of the scattered radiation in a volume of space at the present epoch adds to the list of "numerical coincidences" that seems to suggest we are living in a special epoch[8], a conclusion that although contradicts the Copernican Principle seems to find some justification in the Anthropic Principle[9].

It is interesting to note that the ratio of the self-gravitational energy density of matter to the energy density of the scattered *CMB*

$$\frac{u_{bG}}{u_*} = \frac{4\pi}{5} \frac{G m_p^2}{u_\gamma \sigma_T^2 (1 - Y_p/2)} \tag{9}$$

(where $m_p$ is the mass of the proton and $Y_p$ is the helium mass fraction) is the same in any volume of space, a not very often encounter property when dealing with gravitation, where the sum of the energies of the constituent sub-systems is not equal to the total energy of the system.